\documentclass[twocolumn]{revtex4}
\usepackage{amsmath}
\usepackage{graphicx}
\usepackage{epstopdf}
\usepackage{hyperref}
\usepackage{graphicx,subfigure,xcolor}
\usepackage{braket}
\usepackage{txfonts}
\usepackage{ulem}

\newcommand{\bea}{\begin{eqnarray}}
\newcommand{\eea}{\end{eqnarray}}
\newcommand{\beq}{\begin{equation}}
\newcommand{\eeq}{\end{equation}}
\newcommand{\nn}{\nonumber}

\def\/{\over}

\begin{document}
\title{\bf Radiation-reaction-induced transitions of two maximally entangled atoms  in non-inertial motion}
\author{Wenting Zhou$^{1,2,3}$ and Hongwei Yu$^{3,4}$~\footnote{Corresponding author: hwyu@hunnu.edu.cn}}
\affiliation{$^{1}$ Department of Physics, School of physical science and technology, Ningbo University - Ningbo, Zhejiang 315211, China\\
$^{2}$ Center for Nonlinear Science, Ningbo University - Ningbo, Zhejiang 315211, China\\
$^{3}$ China Key Laboratory of Low Dimensional Quantum Structures and Quantum Control of Ministry of Education, Hunan Normal University, Changsha, Hunan 410081, People¡¯s Republic of China\\
$^{4}$ Department of Physics, Synergetic Innovation Center for Quantum Effects and Applications, Hunan Normal University, Changsha, Hunan 410081, China}

\begin{abstract}
We apply the DDC formalism [proposed by Dalibard, Dupont-Roc and Cohen-Tannoudji] to study the average rate of change of energy of two identical two-level atoms interacting with the vacuum massless scalar field in synchronized motion along stationary trajectories. By separating the contributions of vacuum fluctuations and atomic radiation reaction, we first show  that for the two-atom system initially prepared in the factorizable eigenstates $|g_Ag_B\rangle$ and $|e_Ae_B\rangle$, where $g$ and $e$ represent the ground state and the excited state of a single atom respectively, both  vacuum fluctuations and atomic radiation reaction contribute to the average rate of change of energy of the two-atom system, and the contribution of vacuum fluctuations is independent of the interatomic separation while that of atomic radiation reaction is dependent on it. This is contrary to the existing results in the literature where vacuum fluctuations are interatomic-separation dependent. However, if the two-atom system is initially prepared in the unfactorizable symmetric/antisymmetric entangled state, the average rate of change of energy of the two-atom system is never perturbed by the vacuum fluctuations, but is totally a result of the atomic radiation reaction. We then consider two special cases of motion of the two-atom system which is initially prepared in the symmetric/antisymmetric entangled state, i.e.,
synchronized inertial motion and  synchronized uniform acceleration. In contrast to the average rate of change of energy of a single uniformly accelerated atom, the average rate of change of energy of the uniformly accelerated two-atom system is nonthermal-like. The effects of noninertial motion on the transitions of states of  the two correlated atoms are also discussed.
\end{abstract}
\maketitle

\section{Introduction}

The cause of spontaneous emission, one of the prominent radiative properties of atoms,  has long been a fascinating problem. So far two heuristic pictures-vacuum fluctuations~\cite{Weisskopf} and radiation reaction~\cite{Ackerhalt} or a combination of them~\cite{Milonni75,Milonni88} have been put forward and  the role they play  in spontaneous emission has been widely discussed~\cite{Senitzky73,Milonni73,Ackerhalt74,Milonni75,Milonni88}.
However, there seems to be an ambiguity between the contributions of vacuum fluctuations and radiation reaction as they are crucially dependent on the ordering of commuting atom and field variables. To resolve this uncertainty,  Dalibard, Dupont-Roc and Cohen-Tannoudji (DDC) proposed that a preferred operator ordering should be chosen so that the Hamiltonians of the contributions of vacuum fluctuations and radiation reaction are Hermitian, and thus  respectively possess  independent physical meanings~\cite{DDC82,DDC84}.  Using the DDC formalism,  atomic radiative properties such as the spontaneous emission and the energy shifts~\cite{Audretsch94,Audretsch95,Audretsch952,Passante98} can be well explained, and the spontaneous excitation  of atoms in non-inertial motion is also predicted~\cite{Audretsch94,Audretsch952,Zhu06,Zhou12}.

In recent years, there has been extensive interest in the radiative properties of entangled atoms~\cite{Menezes15,Menezes161,Arias16,Tian16,Liu18,Menezes162,Rizzuto16,Zhou16,Cai18}, as quantum entanglement is a central notion in quantum information and is also crucial for quantum computing. In a recent work,  the authors calculated the response function of two identical atoms in the maximally entangled state interacting with vacuum massless scalar fields, and concluded that the atomic spontaneous transition rates can be enhanced or inhibited depending on the specific entangled state and the interatomic separation, and the presence of boundaries also modifies the transition rates~\cite{Arias16}. Later, by generalizing the DDC formalism~\cite{DDC82,DDC84}, where    the contributions of vacuum fluctuations and atomic radiation reaction to the rate of change of an  observable of the atom are  distinctively separated, to the case of a two-atom system in interaction with the vacuum electromagnetic field, the radiative processes of the two-atom system were studied in the flat Minkowski spacetime~\cite{Menezes15,Menezes161} as well as in the curved Schwarzschild spacetime~\cite{Menezes162},  and the generation and degradation of entanglement of the two-atom system were also analyzed. Following these works, the radiative processes of the same two-atom system in interaction with the vacuum massless scalar field in the de Sitter spacetime~\cite{Liu18} and in interaction with the vacuum electromagnetic field in the cosmic string spacetime~\cite{Cai18} were also investigated. However, as we will show later,  the contributions of vacuum fluctuations are erroneously calculated in Refs.~\cite{Menezes15,Menezes161,Menezes162,Liu18,Cai18}, and as a consequence,  the resulting average rates of change  of energy of the two-atom system  were also incorrect~\footnote{In Ref.~\cite{Cai18}, though the value of the total rate of change of energy of the two-atom system initially prepared in the symmetric/antisymmetric entangled state appears to be correct, the expression of the contribution of vacuum fluctuations [Eq.~(32) of Ref.~\cite{Cai18}] is not.}.

The paper is organized as follows. In section I, we give a detailed derivation of the DDC formalism for study of the average rate of change of energy of a system composed of two identical two-level atoms  in interaction with the vacuum massless scalar field, which are initially prepared in one of the eigenstates. We show that, for the two-atom system initially prepared in the state $|g_Ag_B\rangle$ or $|e_Ae_B\rangle$ where $g$ and $e$ represent the ground state and the excited state of a single atom, both vacuum fluctuations and atomic radiation reaction contribute to the average rate of change of energy of the two-atom system, and furthermore, the contribution of the vacuum fluctuations is independent of the interatomic separation  contrary to the existing results in the literature~\cite{Menezes15,Menezes161,Menezes162,Liu18,Cai18}; while for the two-atom system initially prepared in the symmetric/antisymmetric entangled state, the energy of the two-atom system is never perturbed by the vacuum fluctuations, and the transitions are wholly induced by the atomic radiation reaction. 
In sections III and IV, we use the DDC formalism to calculate the average rate of change of energy of the two-atom system initially prepared in the symmetric/antisymmetric entangled state in two cases: two atoms in synchronized inertial motion and two atoms in synchronized uniform acceleration. By comparing the results in the two cases, we show how the transition processes of the two-atom system are affected by the noninertial motion. We present our conclusions in section V. Throughout the paper, we use the natural units $\hbar=c=1$.

\section{the DDC formalism}
We consider a system of two identical two-level atoms labeled by $A$ and $B$ which are in interaction with the vacuum massless scalar field. The two atoms are assumed to move synchronously and thus the interatomic separation is a constant. We denote the ground state and the excited state of the atoms with energies $-{\omega_0\/2}$ and $+{\omega_0\/2}$ by $|g\rangle$ and $|e\rangle$ respectively, then the Hamiltonian of the two atoms is given by
\beq
H_s=\omega_0R^A_3(\tau)+\omega_0R^B_3(\tau)\label{Hs}
\eeq
with $R_3={1\/2}(|e\rangle\langle e|-|g\rangle\langle g|)$. The Hamiltonian of the scalar field is
\beq
H_F(\tau)=\int d^3k \omega_{\mathbf{k}}a^{\dag}_{\mathbf{k}}a_{\mathbf{k}}{dt\/d\tau}\;.
\eeq
Hereafter, $t$ and $\tau$ represent the coordinate time and the proper time respectively. The interaction between the atoms and the scalar field can be depicted by
\beq
H_I(\tau)=\mu R_2^A(\tau)\phi(x_A(\tau))+\mu R_2^B(\tau)\phi(x_B(\tau))\;,
\eeq
where $\mu$ is the coupling constant which is assumed to be very small, $R_2={i\/2}(R_{-}-R_{+})$ with $R_{-}=|g\rangle\langle e|$ and $R_{+}=|e\rangle\langle g|$, and
\beq
\phi(x)={1\/(2\pi)^{3/2}}\int d^3k {1\/\sqrt{2\omega_{\mathbf{k}}}}\biggl[a_{\mathbf{k}}(t)e^{i\mathbf{k}\cdot\mathbf{x}}+a^{\dag}_{\mathbf{k}}(t)e^{-i\mathbf{k}\cdot\mathbf{x}}\biggr]\label{field-operator}
\eeq
is the scalar field operator with $a_{\mathbf{k}}(t)$ and $a^{\dag}_{\mathbf{k}}(t)$ being the annihilation and creation operators respectively. The total Hamiltonian of the system  is obtained by summing up the above three Hamiltonians:
\bea
H(\tau)&=&\omega_0R^A_3(\tau)+\omega_0R^B_3(\tau)+\int d^3k \omega_{\mathbf{k}}a^{\dag}_{\mathbf{k}}a_{\mathbf{k}}{dt\/d\tau}\nn\\&&
+\mu [R_2^A(\tau)\phi(x_A(\tau))+ R_2^B(\tau)\phi(x_B(\tau))]\;.
\eea

Next, we follow the DDC formalism to calculate the average rate of change of energy of the two-atom system in terms of the contributions of vacuum fluctuations and atomic radiation reaction in the Heisenberg picture.

Starting from the above Hamiltonian, we can derive the following Heisenberg equation of motion for the dynamical variables, $a_{\mathbf{k}}(t(\tau))$, of the field
\bea
{d\/d\tau}a_{\mathbf{k}}(t(\tau))&=&-i\omega a_{\mathbf{k}}(t(\tau)){dt\/d\tau}+i\mu R_2^{A}(\tau)[\phi(x_A(\tau)),a_{\mathbf{k}}(t(\tau))]\nn\\&&
+i\mu R_2^{B}(\tau)[\phi(x_B(\tau)),a_{\mathbf{k}}(t(\tau))]
\label{Heisenberg-equation-field-variable}
\eea
with $[ , ]$ denoting the commutator of two operators, and those of the atoms
\bea
&{d\/d\tau}R^{\xi}_{\pm}(\tau)=i\omega_0[R^{\xi}_3(\tau),R^{\xi}_{\pm}(\tau)]+i\mu[R^{\xi}_2(\tau),R^{\xi}_{\pm}(\tau)]\phi(x_{\xi}(\tau))\;,\\
&{d\/d\tau}R^{\xi}_{3}(\tau)=i\mu[R^{\xi}_2(\tau),R^{\xi}_3(\tau)]\phi(x_{\xi}(\tau))\;,
\label{Heisenberg-equation-atomic-variable}
\eea
where $\xi=A,B$.

Solutions of the above  equations (\ref{Heisenberg-equation-field-variable})-(\ref{Heisenberg-equation-atomic-variable}) can be divided into two parts: the free part which exists even when there is no coupling between the atoms and the field, and the source part which is induced by the interaction between the atoms and the field, i.e.
\bea
a_{\mathbf{k}}(t(\tau))&=&a^{f}_{\mathbf{k}}(t(\tau))+a^{s}_{\mathbf{k}}(t(\tau))\;,\label{expansion-a}\\
R^{\xi}_{\pm}(\tau)&=&R^{\xi f}_{\pm}(\tau)+R^{\xi s}_{\pm}(\tau)\;,\label{expansion-Rpm}\\
R^{\xi}_3(\tau)&=&R^{\xi f}_3(\tau)+R^{\xi s}_3(\tau)\;.\label{expansion-R3}
\eea
Up to the first order of the coupling constant $\mu$, the free part and the source part of the dynamical variables of the field are found to be
\bea
\left\{
  \begin{array}{ll}
    a^{f}_{\mathbf{k}}(t(\tau))=a^{f}_{\mathbf{k}}(t(\tau_0))e^{-i\omega(t(\tau)-t(\tau_0))}\;, \\
    a^{s}_{\mathbf{k}}(t(\tau))=i\mu\sum_{\xi=A,B}\int^{\tau}_{\tau_0}d\tau'R^{\xi f}_2(\tau')[\phi^f(x_{\xi}(\tau')),a^{f}_{\mathbf{k}}(t(\tau))]\;.
  \end{array}
\right.
\eea
Correspondingly, the free part and the source part of the field operator follow
\bea
\phi^f(x(\tau))={1\/(2\pi)^{3/2}}\int {d^3k\/\sqrt{2\omega_{\mathbf{k}}}}\;\biggl[a^f_{\mathbf{k}}(t(\tau))e^{i\mathbf{k}\cdot\mathbf{x}}+a^{\dag f}_{\mathbf{k}}(t(\tau))e^{-i\mathbf{k}\cdot\mathbf{x}}\biggr]\label{free-field-operator}\nn\\
\eea
and
\beq
\phi^s(x(\tau))=i\mu\sum_{\xi=A,B}\int^{\tau}_{\tau_0}d\tau'R^{\xi f}_2(\tau')[\phi^f(x_{\xi}(\tau')),\phi^{f}(x(\tau))]\;.
\eeq

Similarly, the free parts and source parts of the atomic dynamical variables are found to be
\bea
\left\{
  \begin{array}{ll}
    R^{\xi f}_{\pm}(\tau)=R^{\xi}_{\pm}(\tau_0)e^{\pm i\omega_0(\tau-\tau_0)}\;, \\
    R^{\xi s}_{\pm}(\tau)=i\mu\int^{\tau}_{\tau_0}d\tau'[R^{\xi f}_2(\tau'),R^{\xi f}_{\pm}(\tau)]\phi^f(x_{\xi}(\tau'))\;,
  \end{array}
\right.
\label{two parts Rpm}
\eea
and
\bea
\left\{
  \begin{array}{ll}
    R^{\xi f}_3(\tau)=R^{\xi}_3(\tau_0)\;,\\
    R^{\xi s}_3(\tau)=i\mu\int^{\tau}_{\tau_0}d\tau'[R^{\xi f}_2(\tau'),R^{\xi f}_3(\tau)]\phi^f(x_{\xi}(\tau'))\;.
  \end{array}
\right.
\label{two parts R3}
\eea

Now with the operators of the atoms and the fields divided into the free parts and the source parts, we consider the contributions of the vacuum fluctuations and atomic radiation reaction to the average rate of change of energy of the two-atom system. Suppose that the atoms are initially prepared in one of the following states:
\bea
&&|\psi_1\rangle=|g_Ag_B\rangle\;,\nn\\&&
|\psi_2\rangle=|\psi_{\pm}\rangle={1\/\sqrt{2}}(|g_A e_B\rangle\pm|e_Ag_B\rangle)\;,\\&&
|\psi_3\rangle=|e_A e_B\rangle\;.\nn
\eea
These states are eigenstates of the Hamiltoninan of the two-atom system $H_s$[see Eq.~\ref{Hs}] with corresponding energies $-\omega_0,0,\omega_0$, and they form a complete basis.

For atom $A$, the Heisenberg equation of motion with Hamiltonian $H_A(\tau)=\omega_0R_3^A(\tau)$ satisfies
\beq
{d\/d\tau}H_A(\tau)=i\mu\omega_0[R_2^{A}(\tau),R^A_3(\tau)]\phi(x_A(\tau))\;.
\eeq
Replacing the field operator in the above equation with $\phi(x_A(\tau))=\phi^f(x_A(\tau))+\phi^s(x_A(\tau))$ and choosing the symmetric operator ordering for the variables of the atoms and the field as in Refs.~\cite{DDC82,DDC84}, we obtain the operator for the  rate of change  of energy of atom $A$  caused by  the free field, $\phi^f(x(\tau))$, namely the contribution of vacuum fluctuations,

\beq
\biggl({d H_A(\tau)\/d\tau}\biggr)_{vf}={1\/2}i\mu\omega_0\{[R_2^A(\tau),R_3^A(\tau)],\phi^f(x_A(\tau))\}\;,
\eeq
and that by the source field, $\phi^s(x(\tau))$, namely the contribution of atomic radiation reaction,
\beq
\biggl({d H_A(\tau)\/d\tau}\biggr)_{rr}={1\/2}i\mu\omega_0\{[R_2^A(\tau),R_3^A(\tau)],\phi^s(x_A(\tau))\}\;.
\eeq
Hereafter, we denote the anticommutator of two operators by $\{,\}$. By the use of Eqs.~(\ref{expansion-Rpm},\ref{expansion-R3})) and (\ref{two parts Rpm},\ref{two parts R3}) in each of the above two equations, the contributions of vacuum fluctuations and atomic radiation reaction to the second order of the coupling constant can be re-expressed as
\bea
\biggl({d H_A(\tau)\/d\tau}\biggr)_{vf}&=&{1\/2}i\mu\omega_0\{[R_2^{Af}(\tau),R_3^{Af}(\tau)],\phi^f(x_A(\tau))\}\nn\\&
-&{1\/2}\mu^2\omega_0\int^{\tau}_{\tau_0} d\tau'[R_2^{Af}(\tau'),[R_2^{Af}(\tau),R_3^{Af}(\tau)]]\nn\\&&\quad\times\{\phi^f(x_A(\tau)),\phi^f(x_A(\tau'))\}\;,
\eea
\bea
\biggl({d H_A(\tau)\/d\tau}\biggr)_{rr}&=&{1\/2}\mu^2\omega_0\int^{\tau}_{\tau_0} d\tau'\{[[R_2^{Af}(\tau),R_3^{Af}(\tau)],R_2^{Af}(\tau')]\}\nn\\&&\quad\times[\phi^f(x_A(\tau)),\phi^f(x_A(\tau'))]\nn\\
&+&{1\/2}\mu^2\omega_0\int^{\tau}_{\tau_0} d\tau'\{[[R_2^{Af}(\tau),R_3^{Af}(\tau)],R_2^{Bf}(\tau')]\}\nn\\&&\quad\times[\phi^f(x_A(\tau)),\phi^f(x_B(\tau'))]\;.
\eea
Taking the average  of the above two operators over the vacuum state of the field $|0\rangle$, we get
\bea
\biggl\langle{d H_A(\tau)\/d\tau}\biggr\rangle_{vf}&=&i\mu^2\int^{\tau}_{\tau_0} d\tau'C^F(x_A(\tau),x_A(\tau'))\nn\\&&
\times{d\/d\tau}[R_2^{Af}(\tau),R_2^{Af}(\tau')]\;,
\label{average-field-vf}
\eea
and
\bea
&&\quad\biggl\langle{d H_A(\tau)\/d\tau}\biggr\rangle_{rr}=i\mu^2\int^{\tau}_{\tau_0} d\tau'\chi^F(x_A(\tau),x_A(\tau'))\nn\\&&\times{d\/d\tau}\{R_2^{Af}(\tau),R_2^{Af}(\tau')\}+
i\mu^2\int^{\tau}_{\tau_0} d\tau'\chi^F(x_A(\tau),x_B(\tau'))\nn\\&&\times{d\/d\tau}\{R_2^{Af}(\tau),R_2^{Bf}(\tau')\}\;,
\label{average-field-rr}
\eea
in which $\langle\cdots\rangle=\langle 0|\cdots|0\rangle$, and $C^F(x_{\xi}(\tau),x_{\xi'}(\tau'))$ and $\chi^F(x_{\xi}(\tau),x_{\xi'}(\tau'))$ are the symmetric and antisymmetric correlation functions of the field defined as
\bea
C^F(x_{\xi}(\tau),x_{\xi'}(\tau'))&=&{1\/2}\langle0|\{\phi^f(x_{\xi}(\tau)),\phi^f(x_{\xi'}(\tau'))\}|0\rangle\;,\\
\chi^F(x_{\xi}(\tau),x_{\xi'}(\tau'))&=&{1\/2}\langle0|[\phi^f(x_{\xi}(\tau)),\phi^f(x_{\xi'}(\tau'))]|0\rangle\label{definition-chi-f}
\eea
with $\xi,\xi'=A,B$. In obtaining Eqs.~(\ref{average-field-vf}) and (\ref{average-field-rr}), we have used the relation $H_A(\tau)=\omega_0R_3^{Af}(\tau)$ which is accurate to the leading order.

Averaging  Eqs.~(\ref{average-field-vf}) and (\ref{average-field-rr}) over the initial state of the two-atom system, $|\psi_{n}\;\rangle$($n=1,2,3$), we find the contributions of vacuum fluctuations and atomic radiation reaction to the average rate of change of energy of atom $A$:
\bea
\biggl\langle{d H_A(\tau)\/d\tau}\biggr\rangle_{n,vf}=2i\mu^2\int^{\tau}_{\tau_0} d\tau'C^F(x_A(\tau),x_A(\tau')){d\/d\tau}\chi_n^{A}(\tau,\tau')\;,\nn\\
\label{average--vf}
\eea
\bea
\biggl\langle{d H_A(\tau)\/d\tau}\biggr\rangle_{n,rr}&=&2i\mu^2\int^{\tau}_{\tau_0} d\tau'\chi^F(x_A(\tau),x_A(\tau')){d\/d\tau}C_n^{AA}(\tau,\tau')\nn\\
&+&2i\mu^2\int^{\tau}_{\tau_0} d\tau'\chi^F(x_A(\tau),x_B(\tau')){d\/d\tau}C_n^{AB}(\tau,\tau')\;,\nn\\
\label{average--rr}
\eea
in which $\chi_n^{\xi}(\tau,\tau')$ and $C_n^{\xi\xi'}(\tau,\tau')$ are two statistical functions of the atoms defined as
\bea
\chi_n^{\xi}(\tau,\tau')={1\/2}\langle\psi_n|[R_2^{\xi f}(\tau),R_2^{\xi f}(\tau')]|\psi_n\rangle\;,\label{chi-single-atom}\\
C_n^{\xi\xi'}(\tau,\tau')={1\/2}\langle\psi_n|\{R_2^{\xi f}(\tau),R_2^{\xi' f}(\tau')\}|\psi_n\rangle\;.\label{C-atoms-definition}
\eea
The contributions of vacuum fluctuations and atomic radiation reaction to the average rate of change of energy of atom $B$ can be easily obtained by replacing $A$ with $B$ in  Eqs.~(\ref{average--vf}) and (\ref{average--rr}). Thus for the average rate of change of energy of the two-atom system, the contribution of vacuum fluctuations is
\bea
\biggl\langle{d H_s(\tau)\/d\tau}\biggr\rangle_{n,vf}=2i\mu^2\int^{\tau}_{\tau_0} d\tau'C^F(x_A(\tau),x_A(\tau')){d\/d\tau}\chi_n^{A}(\tau,\tau')\nn\\
+2i\mu^2\int^{\tau}_{\tau_0} d\tau'C^F(x_B(\tau),x_B(\tau')){d\/d\tau}\chi_n^{B}(\tau,\tau')\;,\nn\\\label{average--vf-tot}
\eea
and that of the atomic radiation reaction is
\bea
\biggl\langle{d H_s(\tau)\/d\tau}\biggr\rangle_{n,rr}&=&\sum_{\xi,\xi'=A,B}2i\mu^2\int^{\tau}_{\tau_0} d\tau'\chi^F(x_{\xi}(\tau),x_{\xi'}(\tau'))\nn\\&&\quad\quad\quad
\times{d\/d\tau}C_n^{\xi\xi'}(\tau,\tau')\;.
\label{average--rr-tot}
\eea
The expression of the contribution of vacuum fluctuations, Eq.~(\ref{average--vf-tot}), is composed of two terms with each  dependent on only one of the atoms, and it differs from the corresponding expressions derived in Refs.~\cite{Menezes15,Menezes161,Menezes162,Cai18,Liu18}, where the expression of the contribution of vacuum fluctuations is composed of four terms with two of them the same to ours in Eq.~(\ref{average--vf-tot}) as those given in Ref.~\cite{Liu18} where the  scalar field is considered and
two terms  similar to ours as in Refs.~\cite{Menezes15,Menezes161,Menezes162,Cai18} where the electromagnetic field is considered, 
and the other two cross terms which are dependent on both atoms. The two superfluous cross terms originate from the erroneous expressions of the source parts of the atomic dynamical variables. As we have shown in the second lines of Eqs.~(\ref{two parts Rpm}, \ref{two parts R3}), the source parts of the dynamical variables of atom $A$, $R_{\pm}^{As}(\tau)$ and $R_3^{As}(\tau)$, are independent of atom $B$, and vice versa. 
However, in the second line of Eq.~(15) in Ref.~\cite{Menezes161},  the source part of the dynamical variable of atom $A$ is related with atom $B$~\footnote{Though Ref.~\cite{Menezes161} dealt with the interaction between the two-atom system in interaction with the electromagnetic field, the derivations are similar.}, 
and consequently it leads to the erroneous expression of the contribution of vacuum fluctuations. The contribution of atomic radiation reaction, Eq.~(\ref{average--rr-tot}), is composed of four terms with two of them dependent on only one of the atoms while the other two cross-terms dependent on both atoms, and thus it is generally interatomic-separation dependent. This is consistent with what is found in Refs.~\cite{Menezes15,Menezes161,Menezes162,Cai18,Liu18}.

For the two-atom system initially prepared in the factorizable state $|\psi_1\rangle=|g_Ag_B\rangle$ or $|\psi_3\rangle=|e_Ae_B\rangle$, it is easy to deduce from Eq.~(\ref{chi-single-atom}) that
\bea
\chi_1^A(\tau,\tau')&=&\chi_1^B(\tau,\tau')=-{1\/8}(e^{i\omega_0(\tau-\tau')}-e^{-i\omega_0(\tau-\tau')})\;,\\
\chi_3^A(\tau,\tau')&=&\chi_3^B(\tau,\tau')={1\/8}(e^{i\omega_0(\tau-\tau')}-e^{-i\omega_0(\tau-\tau')})\;,
\eea
while for the two-atom system initially prepared in the symmetric/antisymmetric entangled state $|\psi_2\rangle=|\psi_{\pm}\rangle$,
\beq
\chi_2^A(\tau,\tau')=\chi_2^B(\tau,\tau')=0\;.
\eeq
The above statistical functions of the atoms together with the expression of the contribution of vacuum fluctuations [Eq.~(\ref{average--vf})] indicate that the contribution of vacuum fluctuations to the average rate of change of energy of the two-atom system initially prepared in the factorizable state $|g_Ag_B\rangle$ or $|e_Ae_B\rangle$ is generally nonzero and \textit{interatomic-separation independent}, while the contribution of vacuum fluctuations to the average rate of change of energy of the two-atom system initially prepared in the unfactorizable symmetric/antisymmetric entangled state vanishes.

Some comments are now in order as our results are contrary to those in the literature. For example, in Ref.~\cite{Liu18}, the average rate of change of energy of a two-atom system in interaction with the massless scalar field in de Sitter spacetime was calculated and the contribution of vacuum fluctuations to the average rate of change of energy of the atoms initially prepared in all the four states [$|g_Ag_B\rangle$, $|\psi_{\pm}\rangle$ and $|e_Ae_B\rangle$] is incorporated into their Eq.~(24), which is characterized by the interatomic separation dependent factor $f_{12}(\Delta\omega,L/2)$. These interatomic separation dependent terms come from the two cross terms in the expression of the contribution of vacuum fluctuations[see the first line of Eq.~(13) in Ref.~\cite{Liu18}]. However, as we have already pointed out
, these terms are actually non-existent. For the case of the two-atom system initially prepared in the symmetric/antisymmetric entangled state, though the contribution of vacuum fluctuations is expressed in terms of the summation of two terms [corresponding to the upward transition process $|\psi_{\pm}\rangle\rightarrow|g_Ag_B\rangle$ and the downward transition process $|\psi_{\pm}\rangle\rightarrow|e_Ae_B\rangle$] with each of them characterized by $f_{12}(\Delta\omega,L/2)$, the two terms actually sum up to zero. Thus the error in this case doesn't carry on to the final contribution of vacuum fluctuations and neither to the total rate of change of energy. But for the cases of the two-atom system initially prepared in the other eigenstate $|g_Ag_B\rangle$ or $|e_Ae_B\rangle$, the allowed transition process induced by the vacuum fluctuations is $|g_Ag_B\rangle\rightarrow|\psi_{\pm}\rangle$ or $|e_Ae_B\rangle\rightarrow|\psi_{\pm}\rangle$, then only one erroneous term in Eq.~(24) remains. Without the other canceling erroneous term, now, the error leads to an extra erroneous nonzero contribution of vacuum fluctuations, which carries on to the total rate of change of energy of the two-atom system. Similar errors were also made in Refs.~\cite{Menezes15,Menezes161,Menezes162,Cai18} where the average rate of change of energy of the two-atom system in interaction with vacuum electromagnetic fields in various spacetime backgrounds are calculated.

It is worth pointing out that the vanishing contribution of vacuum fluctuations for the two-atom system initially prepared in the symmetric/antisymmetric entangled state is physically understandable. As is demonstrated in Ref.~\cite{Audretsch94}, the vacuum fluctuations tend to excite an atom initially in the ground state, while deexcite an atom initially in the excited state,  and when only the contribution of vacuum fluctuations is taken into account, both excitation and deexcitation occur with equal probability. For the two atoms initially prepared in the symmetric/antisymmetric entangled state, each atom has the probability of $1\/2$ to populate the ground state and the excited state, thus the contribution of vacuum fluctuations comes out to be nullified. Same conclusion can also be drawn  if we view the two atoms as a whole. The symmetric/antisymmetric entangled state belongs to the intermediate state of the two-atom system with $zero-$energy. As vacuum fluctuations are equally capable of deexciting and exciting the two-atom system, the average rate of change of energy of the system due to the upward-transition ($|\psi_{\pm}\rangle\rightarrow|e_Ae_B\rangle$) and the downward-transition ($|\psi_{\pm}\rangle\rightarrow|g_Ag_B\rangle$) sums up to zero.

In the following two sections, we are mainly interested in the transitions of the two-atom system initially prepared in the symmetric/antisymmetric entangled state $|\psi_2\rangle=|\psi_{\pm}\rangle$ in two cases: two atoms in synchronized inertial motion and two atoms in synchronized uniform acceleration. As the vacuum fluctuations do not contribute, the average rate of change of energy of the two-atom system in the symmetrci/antisymmetric entangled state is only ascribed to the contribution of atomic radiation reaction,
\bea
&&\biggl\langle{d H_s(\tau)\/d\tau}\biggr\rangle=\biggl\langle{d H_s(\tau)\/d\tau}\biggr\rangle_{rr}\nn\\&&\quad\quad=\sum_{\xi,\xi'=A,B}2i\mu^2\int^{\tau}_{\tau_0} d\tau'\chi^F(x_{\xi}(\tau),x_{\xi'}(\tau'))
{d\/d\tau}C^{\xi\xi'}(\tau,\tau')\;,\nn\\
\label{average-tot}
\eea
Hereafter, we omit the subscript $n=2$ for simplicity. Let us note here that the resonance interatomic energy of the two-atom system in the symmetric/antisymmetric entangled state is also only ascribed to the atomic radiation reaction and irrelevant to the vacuum fluctuations~\cite{Rizzuto16}.

\section{rate of change of energy of two entangled atoms in synchronized inertial motion}
Suppose that two atoms are in synchronized inertial motion along the same direction, and we choose the Cartesian coordinates to depict their trajectories:
\bea
t_A(\tau)=\gamma\tau\;,\;\; x_A(\tau)=x_0+v\gamma\tau\;,\;\; y_A=0\;,\;\; z_A=0\;,\;\;\label{ineritial-A-trajectory}\\
t_B(\tau')=\gamma\tau',\; x_B(\tau')=x_0+v\gamma\tau',\; y_B=L\;,\;\; z_B=0\;,\;\;\label{ineritial-B-trajectory}
\eea
where $v$ denotes the constant velocity of the atoms and $\gamma=(1-v^2)^{-1/2}$.

According to Eq.~(\ref{average-tot}), to calculate the average rate of change of energy of the two atoms, we should firstly derive the antisymmetric correlation function of the field defined in Eq.~(\ref{definition-chi-f}). For the massless scalar field, we have
\bea
\chi^F(x_{\xi}(\tau),x_{\xi'}(\tau'))={i\/8\pi|\Delta\mathbf{x}|}[\delta(\Delta t+|\Delta\mathbf{x}|)-\delta(\Delta t-|\Delta\mathbf{x}|)]\nn\\\label{chi-f-inertial-trajectories}
\eea
where $\Delta t=t_{\xi}(\tau)-t_{\xi'}(\tau')$ and $|\Delta\mathbf{x}|=|\mathbf{x}_{\xi}(\tau)-\mathbf{x}_{\xi'}(\tau')|$,  
which, for two inertial atoms moving along the trajectories (\ref{ineritial-A-trajectory},\ref{ineritial-B-trajectory}), reduces to
\bea
\chi^F(x_A(\tau),x_A(\tau'))&=&\chi^F(x_B(\tau),x_B(\tau'))\nn\\&=&-{i\/4\pi}{\delta(\Delta\tau)\/\Delta\tau}\;,\label{chi-f-single-atom}\\
\chi^F(x_A(\tau),x_B(\tau'))&=&\chi^F(x_B(\tau),x_A(\tau'))\nn\\&=&{i\/8\pi L}[\delta(\Delta\tau+L)-\delta(\Delta\tau-L)] \label{chi-f-two-atoms}
\eea
with $\Delta\tau=\tau-\tau'$. While for the two atoms prepared in the symmetric/antisymmetric entangled state $|\psi_{\pm}\rangle$, the atomic statistical functions defined in Eq.~(\ref{C-atoms-definition}) are found to be
\bea
C^{\xi\xi'}(\tau,\tau')=\left\{
              \begin{array}{ll}
                \;\;{1\/8}(e^{i\omega_0(\tau-\tau')}+e^{-i\omega_0(\tau-\tau')})\;,\quad\quad\xi=\xi'\;,\\
                \pm{1\/8}(e^{i\omega_0(\tau-\tau')}+e^{-i\omega_0(\tau-\tau')})\;,\quad\quad\xi\neq\xi'\;.
              \end{array}
            \right.
\label{C-atoms}
\eea
In the second line of the above equation, the ``$\pm$'' correspond to $|\psi_{\pm}\rangle$ respectively.

Inserting Eqs.~(\ref{chi-f-single-atom})-(\ref{C-atoms}) into Eq.~(\ref{average-tot}) and doing some simplifications, we obtain the expression of the average rate of change of energy of the two atoms:
\bea
\biggl\langle{d H_s(\tau)\/d\tau}\biggr\rangle&=&{\mu^2i\omega_0\/8\pi}\int^{\tau}_{\tau_0}d\tau'\;
(e^{i\omega_0\Delta\tau}-e^{-i\omega_0\Delta\tau}){\delta(\Delta\tau)\/\Delta\tau}\nn\\&\mp&{\mu^2i\omega_0\/16\pi L}\int^{\tau}_{\tau_0}d\tau'\;
(e^{i\omega_0\Delta\tau}-e^{-i\omega_0\Delta\tau})[\delta(\Delta\tau+L)\nn\\&&-\delta(\Delta\tau-L)]\;.
\eea
Taking the time interval $\Delta\tau$ to be infinitely long, the above integrations can be simplified to
\bea
\biggl\langle{d H_s(\tau)\/d\tau}\biggr\rangle&=&-{\mu^2\omega_0^2\/8\pi}\mp{\mu^2\omega_0\/8\pi}{\sin(\omega_0L)\/L}\;,\label{inertial-result}
\eea
The first term on the right of the above result is exactly equal to the average rate of change of energy of a single inertial excited atom coupled to the massless scalar field [see Eqs.~(42,43) in Ref.~\cite{Audretsch94}]. This consistency is physically understandable. As is shown in Ref.~\cite{Audretsch94}, the atomic radiation reaction gives equal contribution to the average rate of change of an atom in the ground state as well as  in the excited state~\cite{Audretsch94}. Thus for the two-atom system prepared in the symmetric/antisymmetric entangled state, the average rate of change of the atomic energy contains the contributions of radiation reaction to both atoms, which sum up to the first term in Eq.~(\ref{inertial-result}). The second term is characterized by the interatomic separation, and the sign of this term is opposite for the symmetric and the antisymmetric entangled states. It manifests the interference effect of the radiative fields of the two entangled atoms. When $\omega_0L\ll1$, $\langle{d H_s(\tau)\/d\tau}\rangle$ for the symmetric entangled state $|\psi_{+}\rangle$ is almost twice of the average rate of change of energy of a single excited atom in interaction with the vacuum scalar field; while $\langle{d H_s(\tau)\/d\tau}\rangle$ for the antisymmetric entangled state $|\psi_-\rangle$ is almost zero. For a general value of the interatomic separation $L$, this average rate of change of the atomic energy can either be enhanced or weakened for the atoms in both the symmetric and the antisymmetric entangled states, as compared to that of a single excited atom.

\section{rate of change of energy of two entangled atoms in synchronized uniform acceleration}

In this section, we calculate the average rate of change of energy of the two atoms in synchronized uniform acceleration with constant interatomic separation. Suppose that the two atoms are uniformly accelerated along the $x-$direction, and their trajectories are depicted by
\bea
&t_A(\tau)={1\/a}\sinh(a\tau)\;,\; x_A(\tau)={1\/a}\cosh(a\tau)\;,\; y_A=0\;,\; z_A=0\;,\;\label{acceleration-A-trajectory}\nn\\\\
&t_B(\tau')={1\/a}\sinh(a\tau'),\; x_B(\tau')={1\/a}\cosh(a\tau'),\; y_B=L,\; z_B=0\;.\;\label{acceleration-B-trajectory}\nn\\
\eea
Combining the above trajectories with Eq.~(\ref{chi-f-inertial-trajectories}), the antisymmetric correlation functions of the field can be expressed as
\beq
\chi^F(x_{\xi}(\tau),x_{\xi}(\tau'))=-{i\/4\pi}{\delta(\Delta\tau)\/{2\/a}\sinh({a\/2}\Delta\tau)}\;,
\eeq
and
\bea
\chi^F(x_{\xi}(\tau),x_{\xi'}(\tau'))&=&{i\/8\pi L\sqrt{1+{1\/4}a^2L^2}}\biggl[\delta\biggl(\Delta\tau+{2\/a}\sinh^{-1}\biggl({aL\/2}\biggr)\biggr)\nn\\&&
-\delta\biggl(\Delta\tau-{2\/a}\sinh^{-1}\biggl({aL\/2}\biggr)\biggr)\biggr]
\eea
for $\xi\neq\xi'$.
The use of the above correlation functions of the field together with Eq.~(\ref{C-atoms}) in Eq.~(\ref{average-tot}) gives the following average rate of change:
\bea
&&\biggl\langle{d H_s(\tau)\/d\tau}\biggr\rangle={\mu^2i\omega_0a\/16\pi}\int^{\tau}_{\tau_0}d\tau'\;
(e^{i\omega_0\Delta\tau}-e^{-i\omega_0\Delta\tau}){\delta(\Delta\tau)\/\sinh({a\/2}\Delta\tau)}\nn\\&&\quad\quad\quad\mp{\mu^2i\omega_0\/16\pi L\sqrt{1+{1\/4}a^2L^2}}\int^{\tau}_{\tau_0}d\tau'\;
(e^{i\omega_0\Delta\tau}-e^{-i\omega_0\Delta\tau})\nn\\&&\quad\quad\quad
\times\biggl[\delta\biggl(\Delta\tau+{2\/a}\sinh^{-1}\biggl({aL\/2}\biggr)\biggr)-\delta\biggl(\Delta\tau-{2\/a}\sinh^{-1}\biggl({aL\/2}\biggr)\biggr)\biggr]\;.\nn\\
\eea
Further simplifications of the above integrations lead to
\bea
\biggl\langle{d H_s(\tau)\/d\tau}\biggr\rangle=-{\mu^2\omega_0^2\/8\pi}\mp{\mu^2\omega_0\/8\pi}{\sin({2\omega_0\/a}\sinh^{-1}({aL\/2}))\/L\sqrt{1+{1\/4}a^2L^2}}\;.
\eea
This is the total average rate of change of energy of two synchronously uniformly accelerated atoms in the symmetric/antisymmetric entangled state and in interaction with the vacuum massless scalar field. The first term is the same as the corresponding term in the case of two inertial atoms [see the first term in Eq.~(\ref{inertial-result})]; the second term is characterized by the interatomic separation $L$ and the atomic acceleration $a$, and thus it exhibits the interference effects of the radiative fields of the two entangled atoms. Obviously, the interference effects in this case are modulated by the atomic noninertial motion. Comparing this result with the average rate of change of energy of a single uniformly accelerated atom in interaction with vacuum scalar fields [see Eq.~(59) of Ref.~\cite{Audretsch94}], we find a sharp distinction, i.e., for the latter case the average rate of change of the atomic energy is identical to that of a static atom immersed in a thermal bath with temperature $T={a\/2\pi}$; while our result for two atoms correlated by the symmetric/antisymmetric state is nonthermal-like. The cause of the distinction is that the energy of the two atoms in the symmetric/antisymmetric entangled state is never perturbed by the vacuum fluctuations but only affected by the atomic radiation reaction, and only the contribution of vacuum fluctuations exhibits thermal-like behaviors for uniformly accelerated atoms~\cite{Audretsch94,Passante98,Zhou16}; while for a single atom in interaction with the massless scalar field, both vacuum fluctuations and atomic radiation reaction contribute. When $a\rightarrow0$, the above result reduces to that in the case of two inertial atoms, Eq.~(\ref{inertial-result}).

Finally we should stress that the above conclusions are valid for the two-atom system initially prepared in the symmetric/antisymmetric entangled state. If the two atoms are initially prepared in the other factorizable  eigenstates, $|g_Ag_B\rangle$ and $|e_Ae_B\rangle$, as we have previously pointed out in section II, the contributions of vacuum fluctuations to the rate of change of energy of the two-atom system is \textit{no longer zero}, and thus the total rate of change of energy of the two-atom system should be ascribed to both vacuum fluctuations and atomic radiation reaction. Due to the contribution of vacuum fluctuations, thermal-like effects would appear for two uniformly accelerated atoms.

\section{Conclusions}
In this paper, we have applied the DDC formalism in calculating the average variation rate of energy of a two-atom system in interaction with the vacuum massless scalar field. We demonstrated [in contrast to the existing results in the literature] that for the two-atom system initially prepared in the factorizable eigenstate $|g_Ag_B\rangle$ or $|e_Ae_B\rangle$, both  vacuum fluctuations and atomic radiation reaction contribute to the average rate of change of energy of the two-atom system, with the contribution of vacuum fluctuations independent of the interatomic separation and that of atomic radiation reaction dependent of it; while if the two-atom system is initially prepared in the unfactorizable symmetric/antisymmetric entangled state, the average rate of change of energy of the two-atom system can only be ascribed to the contribution of atomic radiation reaction.

We then exploited the DDC formalism to investigate the effect of atomic noninertial motion on the rate of change for the two-atom system initially prepared in the symmetric/antisymmetric entangled state. We calculated the average rate of change of energy of the two-atom system in two cases: two atoms in synchronized inertial motion and two atoms in synchronized uniform acceleration. We find that for the two atoms in synchronized inertial motion, the average rate of change of energy is composed of an interatomic-separation independent term which is the same as the average rate of change of energy of a single excited atom in interaction with the vacuum massless scalar field, and an interatomic-separation dependent term, of which the sign differs when the state of the system changes from symmetric entangled state to antisymmetric and vice versa. For the case of two synchronously uniformly accelerated atoms, the average rate of change of energy of the two-atom system exhibits non-thermal behaviors. Our results indicate that the radiative processes of the two-atom system in the symmetric/antisymmetric entangled state can be effectively manipulated by the atomic noninertial motion.

\begin{acknowledgments}
This work was supported in part by the NSFC under
Grants No. 11690034, No. 11435006, No. 11875172, and No. 11405091;
the Research Program of Ningbo University under Grants
No. XYL18027, the K. C. Wong Magna Fund in Ningbo
University, and the Key Laboratory of Low Dimensional
Quantum Structures and Quantum Control of Ministry of
Education under Grants No. QSQC1801.
\end{acknowledgments}


\baselineskip=16pt

\begin{thebibliography}{26}
\bibitem{Weisskopf} V. F. Weisskopf, Naturwissenschaften {\bf 23}, 41 (1635).
\bibitem{Ackerhalt} J. R. Ackerhalt, P. L. Knight, and J. H. Eberly, Phys. Rev. Lett. {\bf 30}, 456 (1973).
\bibitem{Milonni75} P. W. Milonni and W. A. Smith, Phys. Rev. A {\bf 11}, 814 (1975).
\bibitem{Milonni88} P. W. Milonni, Phys. Scr. {\bf T21}, 102 (1988).
\bibitem{Senitzky73} I. R. Senitzky, Phys. Rev. Lett. {\bf 31}, 955 (1973).
\bibitem{Milonni73} P. W. Milonni, J. R. Ackerhalt, and W. A. Smith, Phys. Rev. Lett. {\bf 31}, 958 (1973).
\bibitem{Ackerhalt74} J. R. Ackerhalt and J. H. Eberly, Phys. Rev. D {\bf 10}, 3350 (1974).
\bibitem{DDC82} J. Dalibard, J. Dupont-Roc and C. Cohen-Tannoudji, J. Phys. France {\bf 43}, 1617 (1982).
\bibitem{DDC84} J. Dalibard, J. Dupont-Roc and C. Cohen-Tannoudji, J. Phys. France {\bf 45}, 637 (1984).
\bibitem{Audretsch94} J. Audretsch and R. M$\ddot{u}$ller, Phys. Rev. A {\bf 50}, 1755 (1994).
\bibitem{Audretsch952} J. Audretsch, R. M$\ddot{u}$ller, and M. Holzmann, Class. Quant. Grav. {\bf 12}, 2927 (1995).
\bibitem{Audretsch95} J. Audretsch and R. M$\ddot{u}$ller, Phys. Rev. A {\bf 52}, 629 (1995).
\bibitem{Passante98} R. Passante, Phys. Rev. A {\bf 57}, 1590 (1998).
\bibitem{Zhou12} W. Zhou and H. Yu, Phys. Rev. A {\bf 86}, 033841 (2012).
\bibitem{Zhu06} Z. Zhu, H. Yu, and S. Lu, Phys. Rev. D {\bf 73}, 107501 (2006).
\bibitem{Menezes15} G. Menezes and N. F. Svaiter, Phys. Rev. A {\bf 92}, 062131 (2015).
\bibitem{Rizzuto16} L. Rizzuto, M. Lattuca, J. Marino, A. Noto, S. Spagnolo, W. Zhou, and R. Passante, Phys. Rev. A {\bf 94}, 012121 (2016).
\bibitem{Zhou16} W. Zhou, L. Rizzuto, and R. Passante, Phys. Rev. D {\bf 94}, 105025 (2016).
\bibitem{Menezes161} G. Menezes and N. F. Svaiter, Phys. Rev. A {\bf 93}, 052117 (2016).
\bibitem{Arias16} E. Arias, J. G. Due$\tilde{n}$as, G. Menezes and N. F. Svaiter, JHEP {\bf 07}(2016)147.
\bibitem{Tian16} Z. Tian, J. Wang, J. Jing, and A. Dragan, Sci. Rep. {\bf 6}, 35222 (2016).
\bibitem{Menezes162} G. Menezes, Phys. Rev. D {\bf 94}, 105008 (2016).
\bibitem{Liu18} X. Liu, Z. Tian, J. Wang, and J. Jing, Phys. Rev. D {\bf 97}, 105030 (2018).
\bibitem{Cai18} H. Cai and Z. Ren, Class. Quant. Grav. {\bf 35}, 025016 (2018).
\end{thebibliography}
\end{document}